\documentclass[sigconf]{acmart}

\usepackage{enumitem}
\usepackage{tikz}
\usetikzlibrary{positioning, arrows.meta}

\usepackage{multirow}
\AtBeginDocument{%
  }

\setcopyright{acmlicensed}
\copyrightyear{2018}
\acmYear{2018}
\acmDOI{XXXXXXX.XXXXXXX}
\acmISBN{978-1-4503-XXXX-X/2018/06}

\usepackage{ulem}
\usepackage[acronym]{glossaries}
\makeglossaries
\glsdisablehyper

\newacronym{SE}{SE}{Software Engineering}
\newacronym{AI4SE}{AI4SE}{Artificial Intelligence for Software Engineering}

\newacronym{NLP}{NLP}{Natural Language Processing}
\newacronym{PLM}{PLM}{Pre-trained Language Model}
\newacronym{PCM}{PCM}{Pre-trained Code Model}

\newacronym{SCPD}{SCPD}{Source Code Plagiarism Detection}
\newacronym{SCPDT}{SCPDT}{Source Code Plagiarism Detection Tool}
\newacronym{SCPC}{SCPC}{Source Code Plagiarism Classification}

\newacronym{NL}{NL}{Natural Language}
\newacronym{PL}{PL}{Programming Language}

\newacronym{NLG}{NLG}{Natural Language Generation}

\newacronym{CEM}{CEM}{Code Evaluation Metric}

\newacronym{LM}{LM}{Language Model}
\newacronym{CodeLLM}{CodeLLM}{Code Large Language Model}
\newacronym{LLM}{LLM}{Large Language Model}
\newacronym{SOTA}{SOTA}{state-of-the-art}

\newacronym{GST}{GST}{Greedy String Tiling}
\newacronym{AST}{AST}{Abstract Syntax Tree}
\newacronym{PDG}{PDG}{Program Dependency Graph}

\newacronym{ROC}{ROC}{Receiver Operating Characteristic}
\newacronym{AUROC}{AUROC}{Area Under the Receiver Operating Characteristic Curve}

\newacronym{PRC}{PRC}{Precision-Recall Curve}
\newacronym{AUPRC}{AUPRC}{Area Under the Precision-Recall Curve}

\newacronym{TPR}{TPR}{True Positive Rate}
\newacronym{FPR}{FPR}{False Positive Rate}
\newacronym{PR}{PR}{Precision-Recall}
\newacronym{AP}{AP}{Average Precision}

\newacronym{TSED}{TSED}{Tree Structured Edit Distance}

\begin{document}

\title{Can Code Evaluation Metrics Detect Code Plagiarism?}

\author{Fahad Ebrahim}
\orcid{0000-0002-3194-8414}
\affiliation{%
  \institution{University of Warwick}
  \city{Coventry}
  \country{United Kingdom}
}
\email{Fahad.Ebrahim@warwick.ac.uk}

\author{Mike Joy}
\orcid{0000-0001-9826-5928}
\affiliation{%
  \institution{University of Warwick}
  \city{Coventry}
  \country{United Kingdom}}
\email{M.S.Joy@warwick.ac.uk}


\begin{abstract}
\gls{SCPD} plays an important role in maintaining fairness and academic integrity in software engineering education. \glspl{CEM} are developed for assessing code generation tasks. However, it remains unclear whether such metrics can reliably detect plagiarism across different levels of modification (L1-L6), increasing in complexity. 

In this paper, we perform a comparative empirical study using two open-source labelled datasets, ConPlag (raw and template-free versions) and IRPlag. We evaluate five \glspl{CEM}, namely CodeBLEU, CrystalBLEU, RUBY, \gls{TSED}, and CodeBERTScore. The performance is evaluated using threshold-free ranking-based measures to assess overall, per dataset, and per-level plagiarism performance. The results are compared against \gls{SOTA} \glspl{SCPDT}, JPlag and Dolos.

Our findings show that without preprocessing, Dolos achieves the highest overall ranking performance, while among the individual metrics, CrystalBLEU, CodeBLEU, and RUBY outperform JPlag. Performance is strongest at L1 and drops from L4 onward, while CrystalBLEU remains competitive on L6. With preprocessing, CrystalBLEU surpasses Dolos overall. Per dataset, Dolos achieved the best ranking on the ConPlag raw dataset, while CrystalBLEU was the best-performing metric on the remaining datasets. At the plagiarism levels, Dolos remains strongest on L4, while CrystalBLEU leads most of the remaining difficult levels. These results indicate that \glspl{CEM} are comparable to dedicated tools in terms of ranking metrics. 
 
\end{abstract}

\begin{CCSXML}
<ccs2012>
   <concept>
       <concept_id>10011007.10011074.10011099</concept_id>
       <concept_desc>Software and its engineering~Software verification and validation</concept_desc>
       <concept_significance>500</concept_significance>
   </concept>
   <concept>
       <concept_id>10010147.10010178</concept_id>
       <concept_desc>Computing methodologies~Artificial intelligence</concept_desc>
       <concept_significance>300</concept_significance>
   </concept>
   <concept>
       <concept_id>10010405.10010481.10010487</concept_id>
       <concept_desc>Applied computing~Education</concept_desc>
       <concept_significance>300</concept_significance>
   </concept>
</ccs2012>
\end{CCSXML}

\ccsdesc[500]{Software and its engineering~Software verification and validation}
\ccsdesc[300]{Computing methodologies~Artificial intelligence}
\ccsdesc[300]{Applied computing~Education}

\keywords{Source Code Plagiarism Detection, Code Evaluation Metrics, Code Similarity, Software Engineering Education}


\maketitle

\section{Introduction}

% Source code similarity 
% - Source code similarity. 
% - application of source code similarity > SCPD. 

Source code similarity refers to the degree to which two programs are alike \cite{zakeri2023systematic}. This similarity can be measured in various ways, ranging from simple text matching to more comprehensive structural and semantic comparisons. These different forms of similarity are used in software engineering applications such as malware detection and vulnerability analysis. Another important application is \gls{SCPD}, which aims to determine whether a program has been copied or reused without proper acknowledgement. In software engineering education, plagiarism can affect fairness and academic integrity \cite{cosma2008towards}. Therefore, reliable \gls{SCPD} methods are essential to ensure fair assessment.

Plagiarism in source code can occur at different levels of modification. Faidhi and Robinson proposed a six-level taxonomy (L1-L6) that can be grouped into three categories \cite{faidhi1987empirical}. The lower levels (L1-L3) involve syntactic changes such as formatting, identifier renaming, and changing the order of statements. The intermediate levels L4 and L5 include structural modifications related to methods, loops, and conditions. The higher level, L6, involves larger changes in semantics, structure, and decision logic. 

Different plagiarism levels require similarity measures that can capture lexical, structural, and semantic aspects of code. \glspl{CEM} were primarily developed for code generation tasks, in which a generated program is compared against a reference implementation for tasks such as code generation, translation, or summarisation. Although they were not developed for plagiarism detection, these metrics can produce numerical similarity scores that can capture different characteristics of source code, such as syntax, structure and semantics. This raises a key question: can \glspl{CEM} detect source code plagiarism across different levels of modifications?

To answer this question, we perform a comparative empirical study using two open-source plagiarism datasets, ConPlag \cite{slobodkin2023towards}, and IRPlag \cite{karnalim_budi_toba_joy_2019}, both labelled with plagiarism levels. We evaluate five \glspl{CEM}: CodeBLEU \cite{ren2020codebleu}, CrystalBLEU \cite{eghbali2022crystalbleu}, RUBY \cite{tran2019does}, \gls{TSED} \cite{song2024revisiting}, and CodeBERTScore \cite{zhou2023codebertscore}. These metrics measure similarity based on token overlap, syntactic structure, tree distance, and embedding representations.

We assess their performance using threshold-free ranking measures, which evaluate how well plagiarised pairs are ranked above non-plagiarised pairs without fixing a decision threshold. We use threshold-free evaluation, as plagiarism detection is highly sensitive to the choice of threshold. Small changes in the threshold can significantly affect the evaluation metrics. The analysis is conducted overall, per dataset, and by plagiarism level to examine how performance changes as modifications become more complex. Finally, we compare the results with \gls{SOTA} \glspl{SCPDT}, JPlag \cite{prechelt2002finding} and Dolos \cite{maertens2022dolos}, to determine how well \glspl{CEM} compare against established plagiarism detection tools. Both JPlag and Dolos preprocess the code by default as part of their algorithms, therefore, we also compare the impact of preprocessing on the performance of \glspl{CEM}. 

This paper makes three main contributions:

\begin{itemize}
    \item We present a systematic evaluation of \glspl{CEM} for source code plagiarism detection. We study multiple datasets and all plagiarism levels (L1-L6). 
    
    \item We show that \glspl{CEM} achieve performance comparable to \glspl{SCPDT} under ranking-based evaluation. We also show that preprocessing is critical. It consistently improves performance and can change the ranking of methods.
    
    \item We show that performance drops from L4 onwards for all methods. This highlights the difficulty of detecting structural and semantic plagiarism. We also show that combining metrics improves ranking accuracy and is a promising direction.
\end{itemize}

%Paper Structure: 
The paper is organised as follows. Section 2 presents the background and related work. Section 3 describes the methodology, including the datasets, the selected \glspl{CEM}, and the evaluation measures. Section 4 presents the results. Section 5 discusses the findings, limitations, threats to validity, and implications. The paper concludes with a summary and an outline of future work.

\section{Background and Related Work}

In this section, we discuss plagiarism levels, \glspl{SCPDT}, and empirical studies involving \glspl{CEM} applied to code similarity tasks. 

\subsection{Plagiarism Levels}

As described by Faidhi and Robinson \cite{faidhi1987empirical}, plagiarism in programming assignments can take place at different levels, ranging from surface edits to heavy logic changes.

\begin{enumerate}[label=L\arabic*)] 
    \item Changing only the appearance of the code, such as editing comments, indentation, or spacing.  
    \item Renaming variables, functions, or classes without changing how the program works.  
    \item Adding, removing, or reordering constants, variables, or functions in the declarations.  
    \item Reorganising functions, for example, by changing their parameters, combining several functions into one, or splitting one function into smaller ones.  
    \item Replacing control structures with equivalent ones, such as using a \texttt{while} loop instead of a \texttt{for} loop, or rewriting conditions differently.  
    \item Changing the internal decision logic by rewriting expressions or conditional statements while keeping the overall behaviour the same.  
\end{enumerate}

As the plagiarism level increases, the changes become more complex and more difficult to detect \cite{alexandra2022material}.

\subsection{Source Code Plagiarism Detection Tools}

% Introductory paragraph about general

In general, \gls{SCPD} methods use different strategies to measure similarity between programs. These approaches range from lexical techniques based on token sequence comparison to structural and semantic methods that analyse representations such as \glspl{AST} \cite{alexandra2022material,zakeri2023systematic}. More recent work often combines multiple signals to capture both surface-level similarity and deeper semantic similarity. Building on these methodological approaches, several systems and tools have been developed to support practical plagiarism detection. 
Several \glspl{SCPDT} have been reviewed and compared in the literature \cite{alexandra2022material,zakeri2023systematic,novak2019source}. 
In this study, we focus on two widely used open-source and actively maintained tools.

JPlag \cite{prechelt2002finding} is a tool that supports languages such as Java, C++, C, and Python. It is based on the \gls{GST} algorithm \cite{wise1993string}, which finds maximal matching substrings between tokenised programs. By comparing token sequences instead of raw text, JPlag reduces the effect of simple formatting changes. However, since it relies on substring matching, it mainly captures lexical and surface-level similarities. JPlag is currently maintained and under active research \cite{sauglam2024obfuscation,sauglam2025mitigating}. 

Dolos \cite{maertens2022dolos} follows a hybrid approach. It combines tokenisation, \gls{AST} parsing, fingerprinting with rolling hash functions, and indexing. By using \glspl{AST}, Dolos captures structural information in addition to token-level similarity. This allows it to detect both lexical and structural similarities and makes it more robust to complex code modifications. Dolos is currently maintained and under active research \cite{maertens2023dolos,maertens2024discovering,maertens2025source}.

\subsection{Code Evaluation Metrics for Code Similarity Tasks}

% Empirical
There have been only a few studies that inspect \glspl{CEM} on code similarity tasks. 

Nikiema et al. systematically examined the robustness of code semantic similarity metrics under controlled transformations for both text and source code \cite{nikiema2025small}. The authors evaluated lexical, embedding-based, structure-aware, and \gls{LLM}-based metrics and showed that several metrics assign high similarity to semantically different code. In particular, embedding-based measures such as CodeBERT \cite{feng2020codebert} and BERTScore \cite{zhang2019bertscore}, which are based on the BERT model \cite{devlin2019bert}, were shown to struggle when small surface changes alter program behaviour. The study also concluded that the choice of distance function, such as cosine versus Euclidean distance, significantly affects semantic interpretation. 

Dristi and Dwyer \cite{dristi2025analyzing} analysed surface bias in reference-based \glspl{CEM}, including CodeBLEU, CrystalBLEU, CodeBERTScore, and CodeScore \cite{dong2025codescore}. They introduced LoCaL, a benchmark dataset of 3117 Python code pairs, labelled with functional similarity using differential fuzzing. The study defined two challenging scenarios, Similar Form Different Semantics and Different Form Similar Semantics, and showed that all four metrics degraded significantly in these cases. CodeBLEU and CodeBERTScore exhibited a strong positive correlation with surface similarity, and CrystalBLEU failed to differentiate when non-equivalent pairs were surface-similar. 

These two works focused on functional and semantic similarity. The evaluation was based on whether two code snippets behave the same or are semantically equivalent. In contrast, plagiarism detection aims to identify copied or derived code, even when modifications change the structure or behaviour. Therefore, this work focuses on plagiarism detection and evaluates how these metrics perform across different levels of plagiarism.

\section{Code Evaluation Metrics as Plagiarism Detectors}
This section presents the overall methodology, datasets, selected \glspl{CEM} and their details, and performance evaluation metrics. 
\subsection{Overall Methodology}

\begin{figure}
    \centering
    \includegraphics[width=1\linewidth]{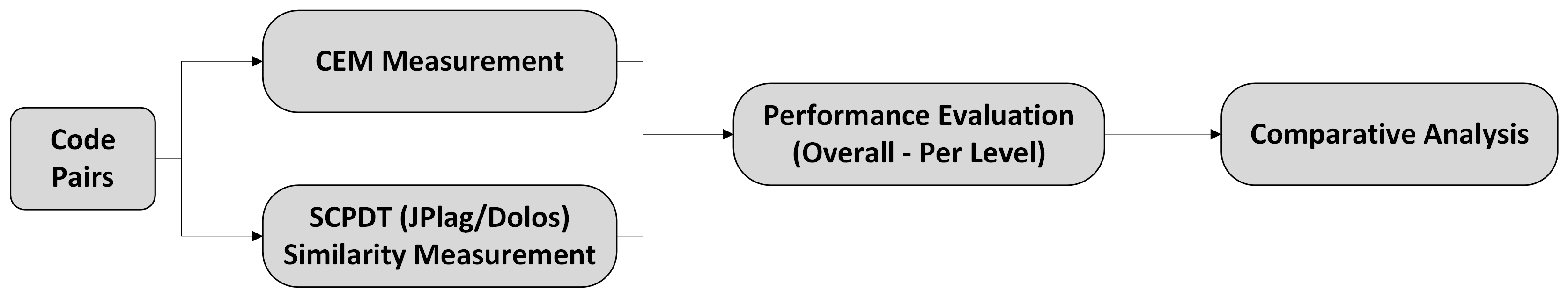}
    \caption{High-Level Methodology.}
    \label{fig:method}
\end{figure}

The high-level methodology used in this paper is presented in Figure \ref{fig:method}. The process begins with the code pairs extracted from the datasets, which form the common input to all similarity estimation steps. We estimate similarity values using the \glspl{CEM} and the tools (JPlag and Dolos). We then evaluate the ranking performance of both the metrics and the tools and conduct a comparative analysis. We also apply preprocessing to the \glspl{CEM} and examine its impact on ranking performance. 

This study aims to answer the following research questions. 

\begin{itemize}
    \item[\textbf{RQ1:}] How well do \glspl{CEM} detect source code plagiarism overall and across different plagiarism levels?

    \item[\textbf{RQ2:}] What is the impact of preprocessing on the performance of \glspl{CEM}, overall and per plagiarism level?
    
    \item[\textbf{RQ3:}] How do \glspl{CEM} compare with the current state-of-the-art plagiarism detection tools in detecting source code plagiarism, overall and per plagiarism level?
    
\end{itemize}

To address RQ3, the main question of the paper, we formulate the following hypothesis \textbf{H}, namely that \glspl{CEM} achieve performance comparable to \glspl{SCPDT} under ranking-based evaluation.

\subsection{Datasets}

There is a scarcity of plagiarism datasets due to their private and sensitive nature. We use two main open-source datasets. The details of both datasets are available in Table \ref{tab:dataset_composition_l1_l6}, in terms of the count of instances, the count of plagiarised and non-plagiarised pairs, along with the count of plagiarised files per level. 

The ConPlag dataset \cite{slobodkin2023towards} is a manually constructed dataset for source code plagiarism in competitive programming contests. It consists of 911 labelled Java solution pairs collected from 21 CodeForces problems, including 251 plagiarised pairs and 660 non-plagiarised pairs. The dataset was built from 4,695 accepted Java submissions and filtered using existing detection tools before final manual labelling. ConPlag is provided in two versions: ConPlag1, a raw version containing the original submissions, and ConPlag2, a template-free version where common contest template code has been manually removed by the authors of the dataset. We label the plagiarism levels (L1-L6) manually on the ConPlag dataset. 

IRPlag \cite{karnalim_budi_toba_joy_2019} is a publicly available dataset designed for evaluating \gls{SCPD}. It consists of 467 Java source code files covering seven introductory programming tasks, including one original solution per task, multiple intentionally plagiarised versions, and independently written non-plagiarised solutions. The plagiarised files were created using different plagiarism levels based on the Faidhi and Robinson taxonomy \cite{faidhi1987empirical}. The plagiarism levels (L1-L6) are already labelled as part of initial annotations. 

The instances in the datasets are relatively long, especially in ConPlag1. Based on the unique code snippets, the average number of tokens per fragment is 1085, 589, and 172 for ConPlag1, ConPlag2, and IRPlag, respectively. This motivates reducing the average token length through preprocessing. The preprocessing steps include removing comments, extra whitespace and new lines, and removing Java import and package statements. After preprocessing, the average number of tokens drops to 958, 483, and 137, corresponding to reductions of 11.72\%, 17.97\%, and 20.34\% for ConPlag1, ConPlag2, and IRPlag, respectively.

For the same unique code snippets, the maximum token counts are 11,909 in ConPlag1, 10,382 in ConPlag2, and 338 in IRPlag for the raw code, compared with 11,677, 10,186, and 283 after preprocessing. For CodeBERTScore, which uses CodeBERT, the maximum token length is 512. The number of unique code fragments exceeding 512 tokens is 743 in ConPlag1 and 443 in ConPlag2, which is reduced to 654 and 268 after preprocessing. None of the IRPlag code fragments exceeds 512 tokens.

\begin{table}[t]
\centering
\caption{Datasets details: counts of non-plagiarised vs.\ plagiarised pairs, number of plagiarised pairs per level (L1-L6), and total number of instances.}
\label{tab:dataset_composition_l1_l6}
\resizebox{\columnwidth}{!}{%
\begin{tabular}{lrrrrrrrrr}
\toprule
Dataset & Non-plag. & Plag. & L1 & L2 & L3 & L4 & L5 & L6 & Total \\
\midrule
ConPlag1 & 660 & 251 & 71 & 4 & 11 & 63 & 67 & 35 & 911 \\
ConPlag2 & 660 & 251 & 71 & 4 & 11 & 63 & 67 & 35 & 911 \\
IRPlag  & 95  & 365 & 61 & 57 & 65 & 60 & 59 & 63 & 460 \\
\midrule
All (Pooled)     & 1415 & 867 & 203 & 65 & 87 & 186 & 193 & 133 & 2282 \\
\bottomrule
\end{tabular}%
}
\end{table}

\subsection{Selected Code Evaluation Metrics}
This section covers the technical details of the selected \glspl{CEM}. 

\subsubsection{BLEU}
Before detailing the \glspl{CEM}, we start with one fundamental metric used for \gls{NLG} evaluation and motivate the code metrics. It is the BLEU metric \cite{papineni2002bleu}, which is an n-gram-based similarity metric originally proposed for machine translation. It measures the overlap between a candidate and a reference text using clipped n-gram precision combined with a brevity penalty. 

Let $C$ be the candidate program and $R$ the reference program.

The clipped precision for n-grams of order $n$ is defined as

\[
p_n =
\frac{
\sum_{\text{n-gram} \in C}
\min(\text{count}_C, \text{count}_R)
}{
\sum_{\text{n-gram} \in C}
\text{count}_C
}.
\]

This term measures how many n-grams in the candidate also appear in the reference. The clipping using $\min(\text{count}_C, \text{count}_R)$ ensures that repeated n-grams in the candidate are not over-counted.

The geometric mean over $N$ n-gram orders is

\[
\text{Prec} =
\exp\left(
\sum_{n=1}^{N}
w_n \log p_n
\right).
\]

This combines precision values for different n-gram lengths (e.g., unigrams, bigrams, and trigrams), so that both short and longer token sequences contribute to the final score.

The brevity penalty is

\[
\text{BP} =
\begin{cases}
1 & \text{if } |C| > |R|, \\
\exp(1 - |R|/|C|) & \text{if } |C| \le |R|.
\end{cases}
\]

The brevity penalty reduces the score if the candidate is much shorter than the reference, preventing very short programs from receiving artificially high similarity values.

The final BLEU score is

\[
\text{BLEU} = \text{BP} \cdot \text{Prec}.
\]

In summary, BLEU ranges from 0 to 1, where higher values indicate greater lexical overlap. It assigns higher values when there is consistent lexical overlap between the candidate and reference across multiple n-gram lengths, while penalising excessively short outputs.

\subsubsection{CodeBLEU}
CodeBLEU extends BLEU by incorporating code-specific information. It combines four components: standard n-gram match, weighted n-gram match, \gls{AST} match, and data-flow match. The final score is a weighted sum of these components. By including structural and data-flow information, CodeBLEU aims to better capture program behaviour compared to pure lexical overlap. 

\[
S_{\text{CodeBLEU}} =
\alpha S_{\text{n-gram}}
+
\beta S_{\text{weighted}}
+
\gamma S_{\text{syntax}}
+
\delta S_{\text{dataflow}},
\]

where

\[
\alpha + \beta + \gamma + \delta = 1.
\]

Here, $S_{\text{n-gram}}$ corresponds to standard BLEU, 
$S_{\text{weighted}}$ assigns higher weights to keywords, 
$S_{\text{syntax}}$ measures \gls{AST} similarity, 
and $S_{\text{dataflow}}$ measures overlap in data dependencies.

\subsubsection{CrystalBLEU}
CrystalBLEU modifies BLEU by removing the most frequent n-grams from the corpus before computing similarity. These frequent n-grams often represent common programming patterns or boilerplate code. By filtering them out, CrystalBLEU reduces inflated similarity caused by shared templates. This makes it more robust to trivial similarities.

Let $G_n(C)$ denote the set of n-grams of order $n$ in program $C$, 
and let $F_n$ denote the set of frequent n-grams in the corpus.

Filtered n-grams are defined as

\[
G_n^{*}(C) = G_n(C) \setminus F_n.
\]

BLEU is then computed using the filtered sets:

\[
\text{CrystalBLEU} =
\text{BLEU}(G_n^{*}(C), G_n^{*}(R)).
\]

\subsubsection{RUBY}
RUBY incorporates structural similarity as it first compares \glspl{PDG} when they can be constructed. If \glspl{PDG} are not available, it falls back to comparing \glspl{AST}. If \glspl{AST} also cannot be constructed, it falls back to a weighted string edit distance between tokenised code. RUBY is defined as

\[
\text{RUBY}(G, R) =
\begin{cases}
\text{GRS}(G, R) & \text{(PDG-based comparison)}, \\
\text{TRS}(G, R) & \text{(AST-based comparison)}, \\
\text{STS}(G, R) & \text{(token-based fallback)}.
\end{cases}
\]

\noindent where $GRS(G,R)$ measures similarity between the \glspl{PDG} of $G$ and $R$,
$TRS(G,R)$ measures similarity between the \glspl{AST} of $G$ and $R$,
and $STS(G,R)$ measures weighted string edit distance between tokenised $G$ and $R$.

\subsubsection{\gls{TSED}}

\gls{TSED} measures structural similarity by computing the minimum cost required to transform one \gls{AST} into another. Let $G_1$ and $G_2$ denote the \glspl{AST} of programs $P_i$ and $P_j$, respectively. The tree edit distance is defined as

\[
\Delta(G_1, G_2) = \min_{\text{ops}} \sum_{i=1}^{n} w(op_i),
\]

where $\text{ops}$ is a sequence of edit operations transforming $G_1$ into $G_2$, and $w(op_i)$ denotes the cost of the $i$th operation, such as insert, delete, or rename. 

To obtain a similarity score, the distance is normalised with respect to the maximum number of nodes between the two trees. The final similarity is computed as

\[
S_{\text{\gls{TSED}}} = \max \left\{ 1 - \frac{\Delta(G_1, G_2)}{\max(\text{Nodes}(G_1), \text{Nodes}(G_2))}, \, 0 \right\}.
\]

\subsubsection{CodeBERTScore}
This metric extends BERTScore by using contextual representations from a pre-trained model, such as CodeBERT, to align tokens between the candidate and reference texts based on their cosine similarity.

\subsection{Performance Evaluation}

We evaluate each \gls{CEM} using threshold-independent measures. In particular, we use the \gls{AUROC} and \gls{AUPRC} \cite{joshi2020machine}. These measures assess how well a metric ranks plagiarised pairs above non-plagiarised pairs without fixing a specific threshold.

Let $S(P_i, P_j) \in [0,1]$ denote the similarity score between two programs. By varying a decision threshold $\tau$, we obtain different classification outcomes. Rather than selecting a single threshold, \gls{AUROC} and \gls{AUPRC} summarise performance across all possible thresholds.

The \gls{ROC} curve plots the \gls{TPR} against the \gls{FPR} at different threshold values. The \gls{TPR} and \gls{FPR} are defined as

\[
\text{TPR} = \frac{\text{TP}}{\text{TP} + \text{FN}}=\text{Recall },\text{ FPR} = \frac{\text{FP}}{\text{FP} + \text{TN}}
\]

\gls{AUROC} measures the probability that a randomly selected plagiarised pair receives a higher similarity score than a randomly selected non-plagiarised pair. A value of 0.5 indicates random ranking, while a value of 1.0 indicates perfect separation. \gls{AUROC} reflects the overall ranking ability of a metric. It is suitable when we are interested in separability between plagiarised and non-plagiarised pairs across all thresholds.

The \gls{PRC} plots Precision against Recall (\gls{TPR}) at different thresholds, with Precision defined as

\[
\text{Precision} = \frac{\text{TP}}{\text{TP} + \text{FP}}
\]
The area under the curve of \gls{PR} corresponds to the \gls{AP}, which is equal to the following

\[
\text{AP} \;=\; \sum_{n=1}^{N} \left(R_{n} - R_{n-1}\right) \cdot P_{n}
,\]

where \(P_n\) is the precision at the \(n\)-th threshold, \(R_n\) is the recall at the \(n\)-th threshold, and \(R_0 = 0\).

\gls{AUPRC} focuses on the trade-off between detecting plagiarised pairs and avoiding false positives. In many plagiarism-detection datasets, non-plagiarised pairs constitute the majority of instances. Therefore, \gls{AUPRC} provides additional insight into practical performance, particularly under class imbalance and is more suitable for top-k ranking \cite{mcdermott2024closer}. A higher \gls{AUPRC} indicates that a metric maintains precision as recall increases. For this reason, both \gls{AUROC} and \gls{AUPRC} are reported to provide a complementary evaluation of ranking performance. 

In addition to overall \gls{AUROC} and \gls{AUPRC} values, we compute these measures separately for each plagiarism level $L_k$. This allows us to examine how metric performance changes with structural and semantic modifications.

\section{Results}

In this section, we present the results of all \glspl{CEM} in addition to JPlag and Dolos. The results are reported for the pooled dataset, for each dataset, and for each plagiarism level (L1–L6). The section also presents the impact of preprocessing on \gls{CEM}. 
For all results, we use bootstrapping \cite{diciccio1996bootstrap} with 10,000 resamples to estimate 95\% confidence intervals for the performance evaluation metrics, following the recommendation of the authors of the ConPlag dataset \cite{slobodkin2023towards}. Bootstrapped 95\% confidence intervals for the pooled raw and preprocessed results are reported in Appendix \ref{app:pooled_confidence_intervals}. We use the default settings for all metrics and tools. For readability, the tables report only the mean, while the values cited in the text include their 95\% confidence intervals, and the full interval tables remain in the appendices and the GitHub repository. 
To test the hypothesis \textbf{H}, we perform paired bootstrap testing as recommended by Dror et al. \cite{dror2018hitchhiker}, as the datasets are small in size and imbalanced.

\subsection{Raw Results}
\begin{figure}
    \centering
    \includegraphics[width=1\linewidth]{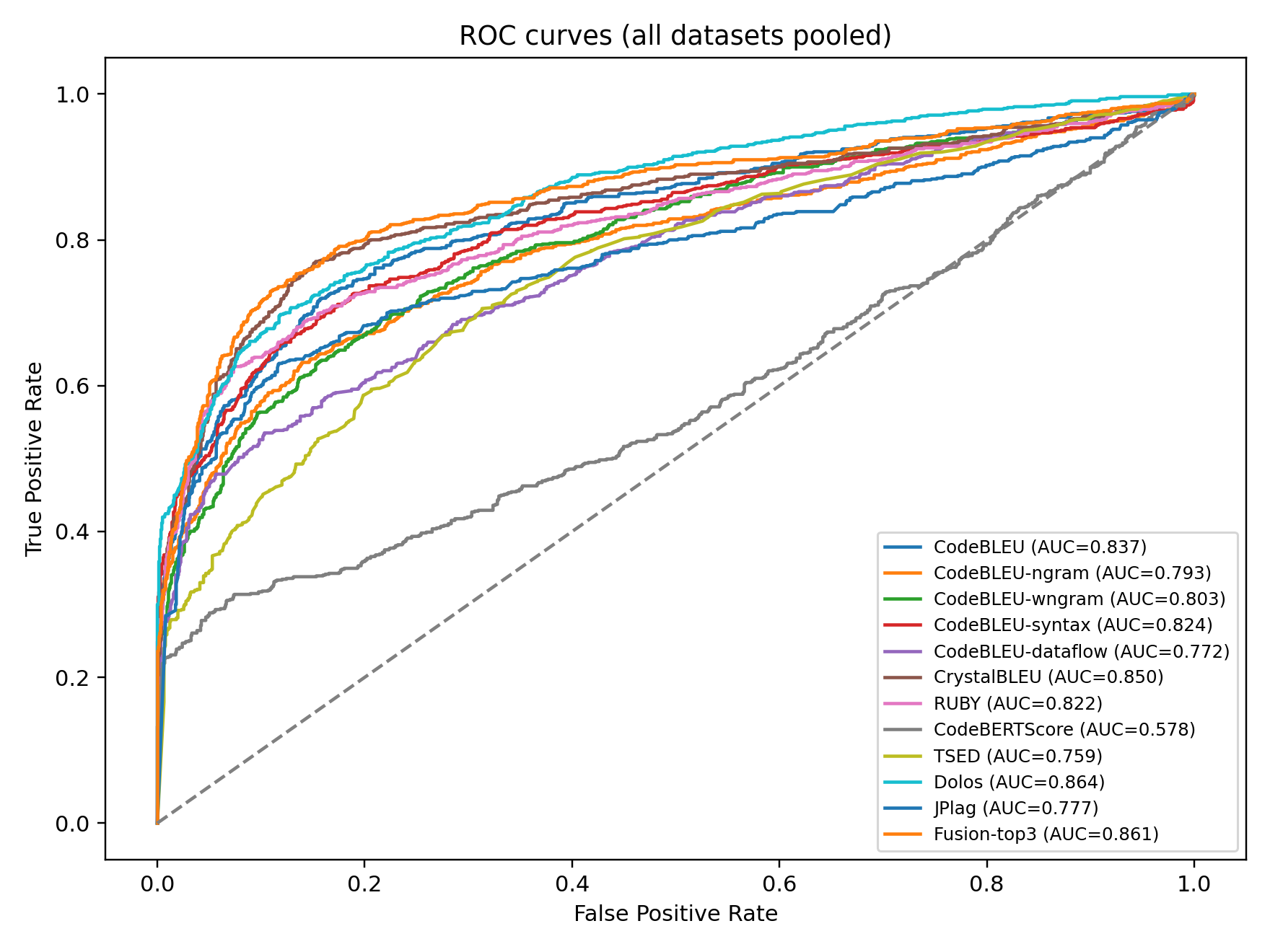}
    \caption{\gls{ROC} curve across the pooled full datasets. }
    \label{fig:raw_roc_curve}
\end{figure}

%Add CodeBLEU and RUBY to the results
We start with the results across the pooled datasets on the raw code snippets. The \gls{ROC} curves presented in Figure \ref{fig:raw_roc_curve} show that among the individual \glspl{CEM}, CrystalBLEU is the best metric with an \gls{AUROC} of 0.850 (95\% CI [0.832, 0.866]), followed by CodeBLEU at 0.837 (95\% CI [0.821, 0.854]) and RUBY at 0.822 (95\% CI [0.804, 0.842]). We also report FusionTop3 as the unweighted average of the three metrics that showed consistently strong performance across datasets (CrystalBLEU, CodeBLEU, and RUBY). These three metrics were in the same range and achieved aligned scores, with a clear gap separating them from the other metrics. This aggregation is exploratory and does not involve parameter tuning or cross-validation. The fusion aims to assess whether combining complementary similarity metrics improves ranking performance without additional tuning. FusionTop3 reaches an \gls{AUROC} of 0.861 (95\% CI [0.845, 0.878]), exceeding all individual metrics and nearly matching Dolos, which attains the highest pooled raw \gls{AUROC} at 0.864 (95\% CI [0.848, 0.879]). JPlag follows at 0.777 (95\% CI [0.753, 0.798]), ahead of \gls{TSED} and CodeBERTScore. %Explain the curve later
The \gls{ROC} curves show how well each method separates plagiarised and non-plagiarised pairs across different thresholds. For stronger methods, the curves rise quickly toward the top-left corner, which means they rank most plagiarised pairs above non-plagiarised ones. As plagiarism levels become more complex, the curves move closer to the diagonal line, indicating weaker separation. This visual pattern matches the decline observed from L4 onwards in the level-wise results.

In addition to the overall CodeBLEU score, we report the performance of its individual components: n-gram match (CB-Ngram), weighted n-gram match (CB-Wngram), syntax match (CB-Syntax), and data-flow match (CB-Dataflow). Across datasets, the syntax component consistently achieves higher ranking values than the other components. For example, in ConPlag2 (raw), CB-Syntax reaches an AUROC of 0.893 (95\% CI [0.870, 0.918]), compared to 0.765 (95\% CI [0.728, 0.801]) for CB-Ngram and 0.702 (95\% CI [0.665, 0.746]) for CB-Dataflow. Similar patterns are observed in other datasets. This indicates that structural similarity contributes more strongly than lexical overlap alone in separating plagiarised and non-plagiarised pairs in this setting. 

\begin{figure}
    \centering
    \includegraphics[width=1\linewidth]{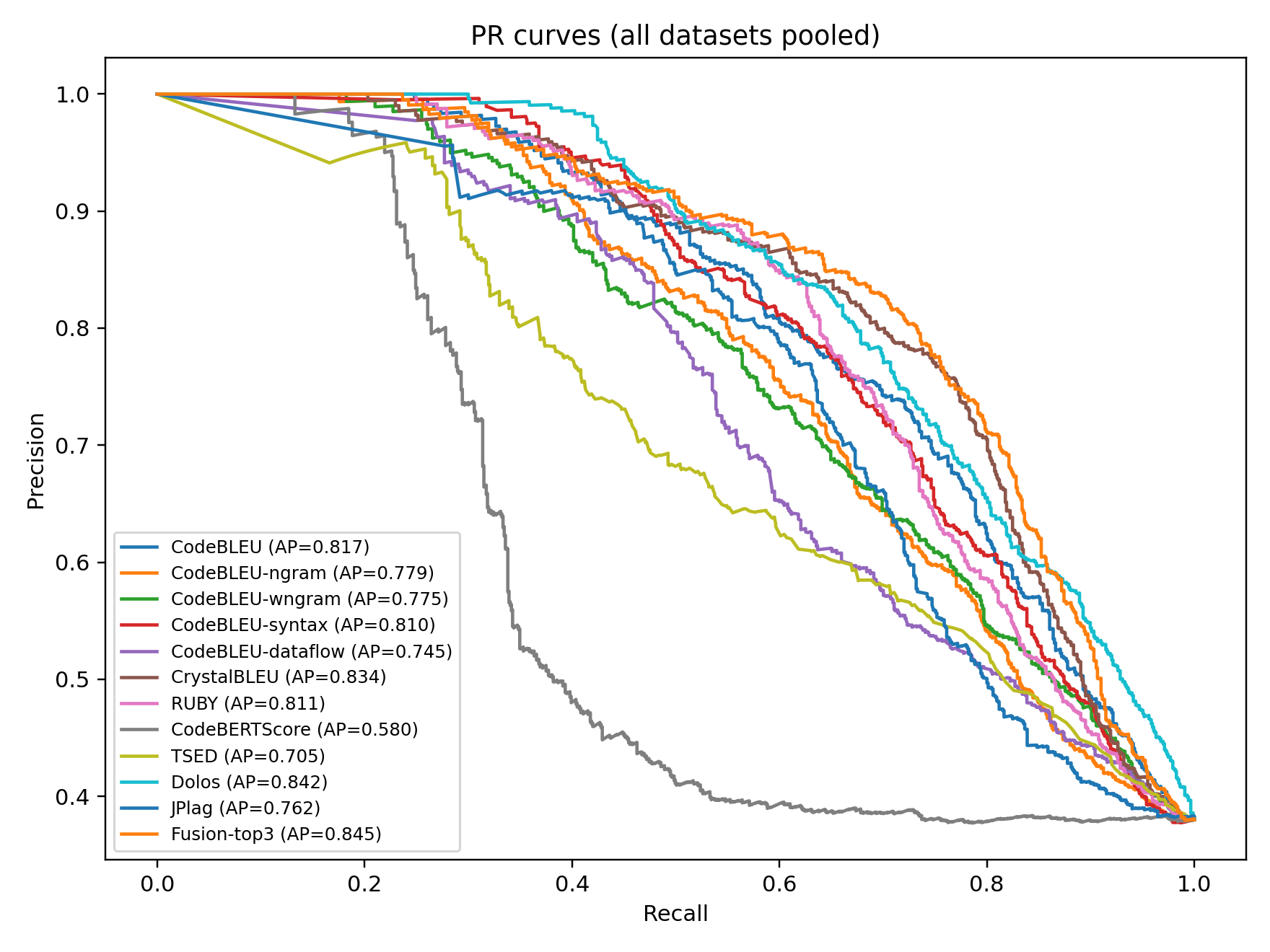}
    \caption{PR curve across the pooled full datasets. }
    \label{fig:raw_pr_curve}
\end{figure}

The \gls{PR} curves for the pooled datasets are available in Figure \ref{fig:raw_pr_curve}. Here, FusionTop3 has the best \gls{AP} value of 0.845 (95\% CI [0.824, 0.861]), followed by Dolos at 0.842 (95\% CI [0.823, 0.858]). Among the individual metrics, CrystalBLEU remains the strongest at 0.834 (95\% CI [0.813, 0.854]), while JPlag follows at 0.762 (95\% CI [0.736, 0.787]). The \gls{PR} curves provide a different view of performance. They show how precision changes as recall increases. In some datasets, precision drops more quickly as recall grows, meaning that detecting more plagiarised pairs also increases false positives.

% Preprocessing Full

\subsection{Impact of Preprocessing on Code Evaluation Metrics}

\begin{figure}
    \centering
    \includegraphics[width=1\linewidth]{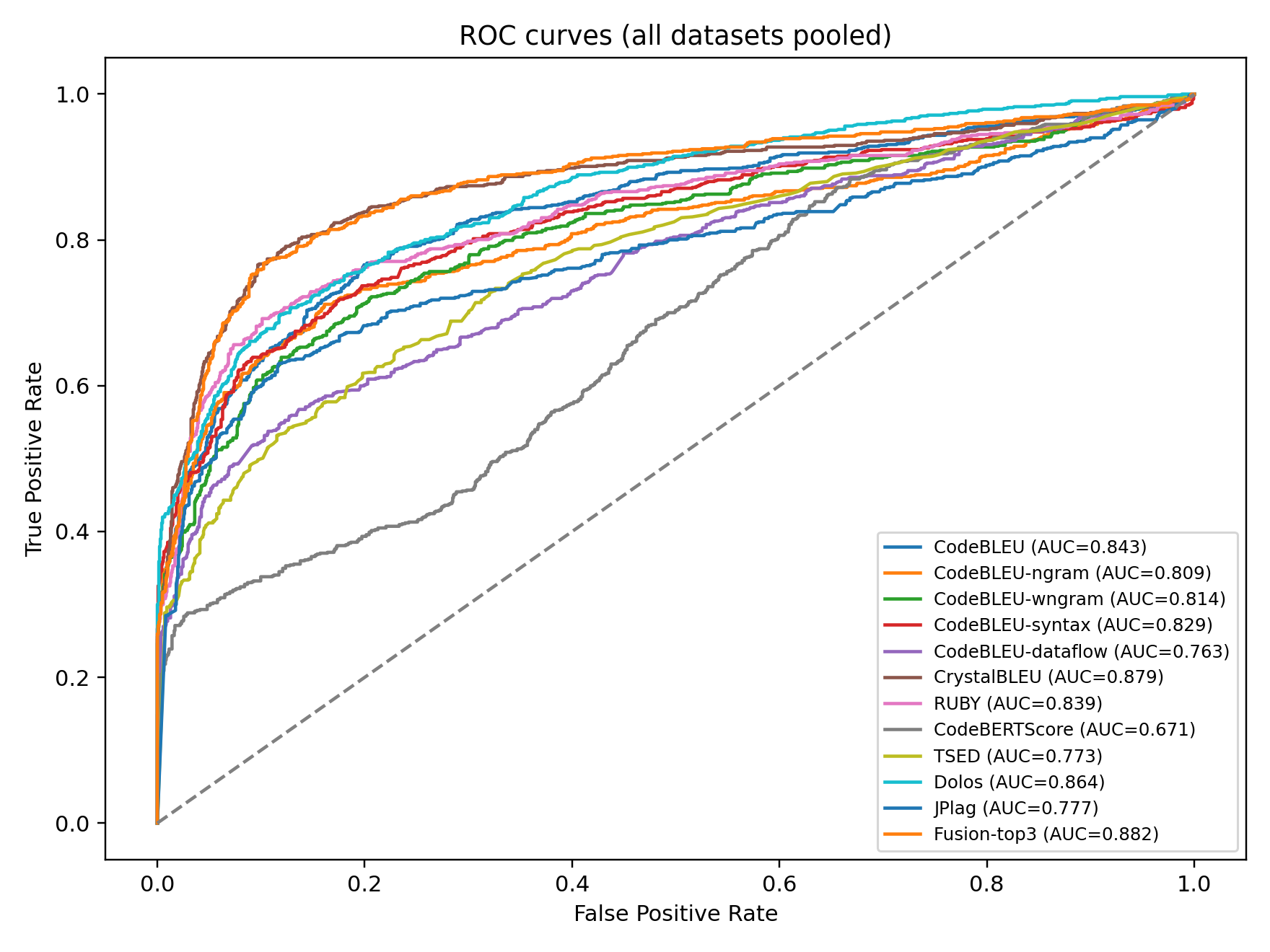}
    \caption{\gls{ROC} curve across the pooled full datasets with preprocessing. }
    \label{fig:pp_roc_curve}
\end{figure}

\begin{figure}
    \centering
    \includegraphics[width=1\linewidth]{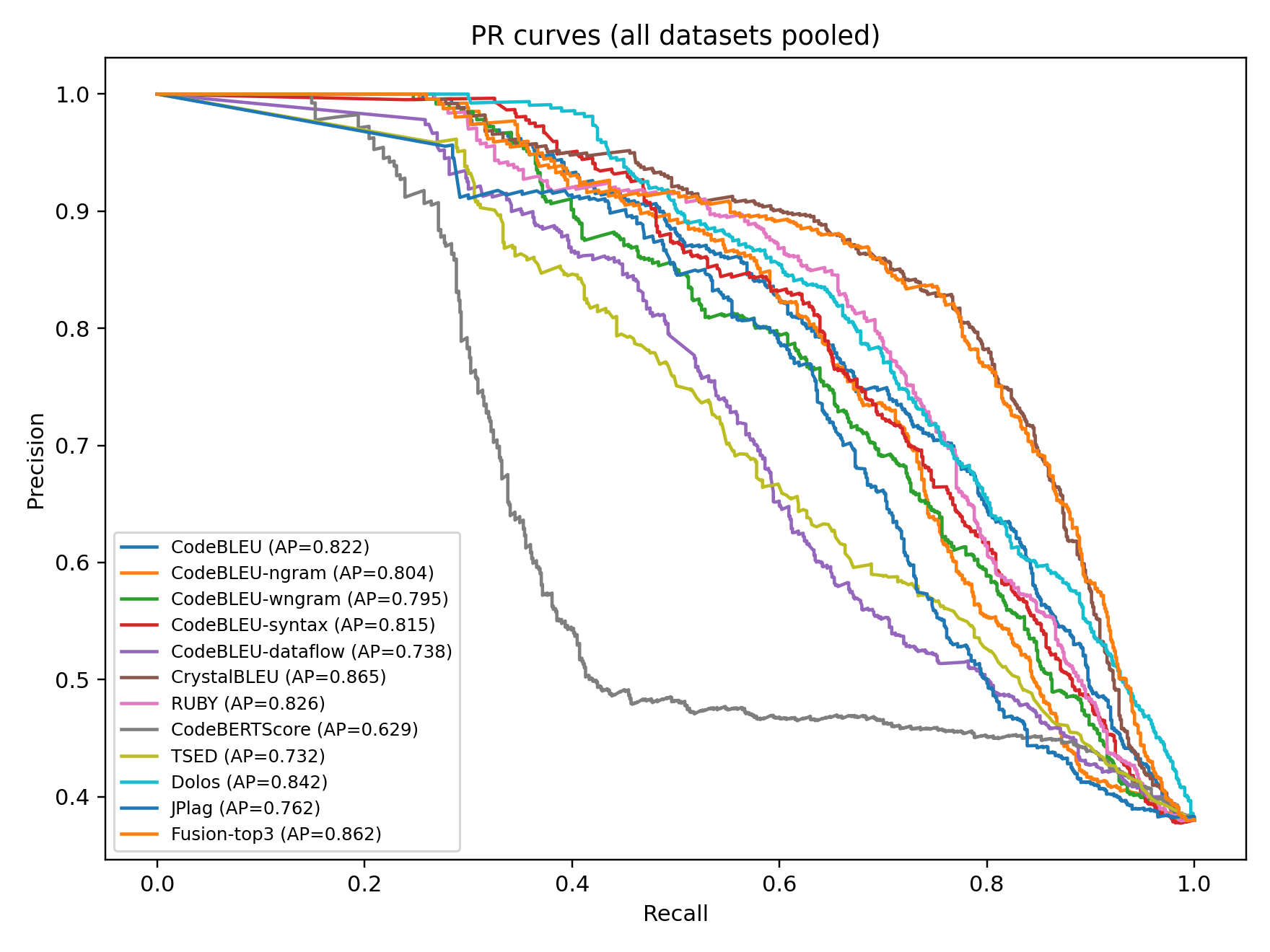}
    \caption{PR curve across the pooled full datasets with preprocessing.}
    \label{fig:pp_pr_curve}
\end{figure}

With preprocessing, the \gls{ROC} curve is available at Figure \ref{fig:pp_roc_curve}. Overall, applying preprocessing improves the performance of most \glspl{CEM}. This leads FusionTop3 and CrystalBLEU to overtake Dolos, with \gls{AUROC} values of 0.882 (95\% CI [0.866, 0.897]) and 0.879 (95\% CI [0.862, 0.895]), respectively, compared with 0.864 (95\% CI [0.848, 0.879]) for Dolos. The remaining metrics also improve, while JPlag stays unchanged at 0.777 (95\% CI [0.753, 0.798]). For CodeBLEU, the relative ordering of the components remains largely unchanged, with the syntax component maintaining higher values than the lexical and data-flow components. % Explain the curve later. 
In terms of the \gls{PR} curve, shown in Figure \ref{fig:pp_pr_curve}, preprocessing also increases most of the \glspl{CEM} performance values. CrystalBLEU is the best method by \gls{AP} with a value of 0.865 (95\% CI [0.848, 0.885]), followed by FusionTop3 at 0.862 (95\% CI [0.841, 0.878]) and then Dolos at 0.842 (95\% CI [0.823, 0.858]). % Explain the curve later. 
So, overall preprocessing increased the performance metrics of most \glspl{CEM}, leading FusionTop3 and CrystalBLEU to surpass the \gls{SOTA} \glspl{SCPDT}, JPlag and Dolos, on the pooled results. 

Paired bootstrap comparisons for the pooled results are reported in Appendix \ref{app:paired_bootstrap_differences}. In the raw setting, the top metrics all outperform JPlag, while only FusionTop3 remains statistically indistinguishable from Dolos. 
After preprocessing, FusionTop3 and CrystalBLEU achieve higher AUROC values than Dolos, whereas CodeBLEU and RUBY remain below Dolos but still well above JPlag. These results support the hypothesis, showing that \glspl{CEM} achieve performance comparable to \glspl{SCPDT} under certain settings.

% Per Dataset
\subsection{Results Per Dataset}

\begin{table*}[t]
\centering
\caption{\gls{AUROC} and \gls{AP} comparison across datasets: Raw (left) vs.\ Preprocessed (right). Bold indicates improvement after preprocessing. Underline indicates the highest value per column within each block.}
\label{tab:auc_ap_raw_vs_pp}
\resizebox{\textwidth}{!}{
\begin{tabular}{lcccccc|cccccc}
\toprule
& \multicolumn{6}{c|}{Raw}
& \multicolumn{6}{c}{Preprocessed} \\
\cmidrule(lr){2-7} \cmidrule(lr){8-13}
& \multicolumn{2}{c}{ConPlag1}
& \multicolumn{2}{c}{ConPlag2}
& \multicolumn{2}{c|}{IRPlag}
& \multicolumn{2}{c}{ConPlag1}
& \multicolumn{2}{c}{ConPlag2}
& \multicolumn{2}{c}{IRPlag} \\
\cmidrule(lr){2-3} \cmidrule(lr){4-5} \cmidrule(lr){6-7}
\cmidrule(lr){8-9} \cmidrule(lr){10-11} \cmidrule(lr){12-13}
Metric 
& AUROC & AP & AUROC & AP & AUROC & AP
& AUROC & AP & AUROC & AP & AUROC & AP \\
\midrule

CodeBLEU
& 0.752 & 0.679 & 0.839 & 0.764 & 0.705 & 0.908
& 0.750 & 0.674 & 0.834 & \textbf{0.764} & 0.692 & 0.906 \\

CB-Dataflow
& 0.693 & 0.597 & 0.702 & 0.611 & 0.660 & 0.877
& 0.685 & \textbf{0.598} & 0.692 & 0.606 & 0.628 & 0.869 \\

CB-Ngram
& 0.689 & 0.635 & 0.765 & 0.694 & 0.658 & 0.897
& 0.684 & 0.629 & 0.756 & \textbf{0.698} & \textbf{0.718} & \underline{\textbf{0.914}} \\

CB-Syntax
& 0.801 & 0.746 & 0.893 & 0.832 & 0.705 & 0.905
& \textbf{0.802} & \textbf{0.747} & \textbf{0.895} & \textbf{0.837} & \textbf{0.706} & \textbf{0.906} \\

CB-Wngram
& 0.719 & 0.647 & 0.782 & 0.708 & 0.651 & 0.892
& 0.712 & 0.641 & 0.773 & \textbf{0.709} & \textbf{0.655} & \textbf{0.897} \\

\midrule

CrystalBLEU
& 0.770 & 0.725 & 0.887 & 0.845 & 0.659 & 0.882
& \textbf{0.779} & \textbf{0.731} & \underline{\textbf{0.942}} & \underline{\textbf{0.897}} & \underline{\textbf{0.722}} & \textbf{0.910} \\

RUBY
& 0.758 & 0.710 & 0.848 & 0.812 & 0.668 & 0.888
& \textbf{0.764} & \textbf{0.715} & \textbf{0.890} & \textbf{0.858} & 0.646 & \textbf{0.890} \\

\gls{TSED}
& 0.715 & 0.606 & 0.723 & 0.590 & 0.564 & 0.866
& \textbf{0.716} & \textbf{0.639} & \textbf{0.746} & \textbf{0.660} & 0.531 & 0.835 \\

CodeBERTScore
& 0.722 & 0.651 & 0.749 & 0.673 & 0.594 & 0.859
& \textbf{0.732} & \textbf{0.663} & \textbf{0.754} & \textbf{0.677} & \textbf{0.657} & \textbf{0.887} \\

\midrule

FusionTop3
& 0.780 & 0.729 & 0.894 & 0.844 & 0.687 & 0.895
& \textbf{0.786} & 0.729 & \textbf{0.935} & \textbf{0.885} & \textbf{0.700} & \textbf{0.903} \\

\midrule

JPlag
& 0.778 & 0.716 & \underline{0.929} & \underline{0.883} & 0.557 & 0.836
& 0.778 & 0.716 & 0.929 & 0.883 & 0.557 & 0.836 \\

Dolos
& \underline{0.840} & \underline{0.770} & 0.916 & 0.860 & \underline{0.717} & \underline{0.913}
& \underline{0.840} & \underline{0.770} & 0.916 & 0.860 & 0.717 & 0.913 \\

\bottomrule
\end{tabular}
}
\end{table*}

We proceed with the results per dataset, which are available in Table \ref{tab:auc_ap_raw_vs_pp} for both \gls{AUROC} and \gls{AP} performance values for raw and preprocessed code pairs. For the raw code fragments, Dolos tops ConPlag1 with AUROC/AP 0.840/0.770 (95\% CIs [0.808, 0.868] / [0.728, 0.819]), JPlag tops ConPlag2 with 0.929/0.883 (95\% CIs [0.907, 0.947] / [0.855, 0.915]), and Dolos also tops IRPlag with 0.717/0.913 (95\% CIs [0.661, 0.764] / [0.889, 0.937]). FusionTop3 is competitive on the two ConPlag datasets, while CrystalBLEU remains the strongest individual metric on ConPlag1 and ConPlag2. With preprocessing, Dolos remains the best method on ConPlag1 at 0.840/0.770 (95\% CIs [0.808, 0.868] / [0.728, 0.819]), CrystalBLEU becomes the best method on ConPlag2 at 0.942/0.897 (95\% CIs [0.925, 0.957] / [0.866, 0.921]), and on IRPlag CrystalBLEU attains the highest \gls{AUROC} at 0.722 (95\% CI [0.665, 0.773]) while CodeBLEU n-gram attains the highest \gls{AP} at 0.914 (95\% CI [0.886, 0.937]). 
An important observation here is that the class distribution had an impact on the metrics. IRPlag has more plagiarised pairs (365) than non-plagiarised pairs (95), while ConPlag1 and ConPlag2 have more non-plagiarised pairs (660) than plagiarised pairs (251). This led to having high \gls{AP} and lower \gls{AUROC} in IRPlag compared to ConPlag. 

\subsection{Results Per Plagiarism Level}

\begin{table*}[t]
\centering
\caption{Pooled overall performance per plagiarism level (L1-L6): Raw (left) vs.\ Preprocessed (right). Metrics reported as AUROC and AP. Bold indicates improvement after preprocessing. Underline indicates the highest value per level within each block.}
\label{tab:pooled_level_auc_ap_raw_vs_pp}
\resizebox{\textwidth}{!}{%
\begin{tabular}{llcccccc|cccccc}
\toprule
& & \multicolumn{6}{c|}{Raw} & \multicolumn{6}{c}{Preprocessed} \\
\cmidrule(lr){3-8} \cmidrule(lr){9-14}
Metric & Score & L1 & L2 & L3 & L4 & L5 & L6 & L1 & L2 & L3 & L4 & L5 & L6 \\
\midrule
CodeBLEU & AUROC & \underline{1.000} & 0.974 & 0.948 & 0.783 & 0.699 & 0.724 & \underline{\textbf{1.000}} & \textbf{0.983} & \textbf{0.958} & 0.775 & \textbf{0.713} & \textbf{0.743} \\
 & AP & \underline{0.998} & 0.792 & 0.630 & 0.379 & 0.268 & 0.209 & \underline{\textbf{1.000}} & \textbf{0.832} & \textbf{0.660} & 0.361 & \textbf{0.269} & 0.203 \\
CB-DF & AUROC & 0.989 & 0.954 & 0.870 & 0.662 & 0.611 & 0.673 & \textbf{0.992} & 0.948 & \textbf{0.900} & 0.640 & 0.606 & 0.635 \\
 & AP & 0.950 & 0.672 & 0.378 & 0.224 & 0.191 & 0.179 & \textbf{0.958} & 0.658 & \textbf{0.393} & 0.213 & 0.181 & 0.142 \\
CB-Ngram & AUROC & 0.992 & 0.911 & 0.912 & 0.709 & 0.630 & 0.708 & \textbf{1.000} & \textbf{0.969} & \textbf{0.927} & 0.708 & \textbf{0.642} & \textbf{0.747} \\
 & AP & 0.968 & 0.533 & 0.534 & 0.376 & 0.262 & 0.224 & \textbf{0.998} & \textbf{0.720} & \textbf{0.638} & 0.366 & \textbf{0.264} & \textbf{0.231} \\
CB-Syntax & AUROC & 0.999 & \underline{0.983} & 0.931 & 0.787 & 0.711 & 0.626 & \textbf{0.999} & 0.986 & 0.934 & \textbf{0.790} & \textbf{0.723} & \textbf{0.630} \\
 & AP & 0.991 & \underline{0.851} & 0.632 & 0.419 & 0.317 & 0.154 & \textbf{0.991} & \underline{\textbf{0.874}} & \textbf{0.658} & \textbf{0.421} & \textbf{0.326} & 0.152 \\
CB-WN & AUROC & 0.995 & 0.912 & 0.899 & 0.742 & 0.639 & 0.714 & \textbf{1.000} & \textbf{0.963} & \textbf{0.911} & 0.733 & \textbf{0.651} & \textbf{0.742} \\
 & AP & 0.977 & 0.469 & 0.461 & 0.332 & 0.234 & 0.189 & \textbf{0.998} & \textbf{0.643} & \textbf{0.578} & \textbf{0.336} & \textbf{0.243} & \textbf{0.203} \\
\midrule
CrystalBLEU & AUROC & 0.994 & 0.930 & 0.955 & 0.811 & 0.745 & 0.727 & \textbf{1.000} & \textbf{0.986} & \textbf{0.976} & \textbf{0.834} & \underline{\textbf{0.794}} & \textbf{0.767} \\
 & AP & 0.973 & 0.583 & 0.679 & 0.483 & 0.373 & \underline{0.233} & \textbf{0.997} & \textbf{0.774} & \underline{\textbf{0.779}} & \textbf{0.485} & \underline{\textbf{0.404}} & \underline{\textbf{0.254}} \\
RUBY & AUROC & 0.995 & 0.955 & 0.941 & 0.766 & 0.711 & 0.651 & \textbf{0.999} & \textbf{0.987} & \textbf{0.946} & \textbf{0.776} & \textbf{0.745} & \textbf{0.677} \\
 & AP & 0.978 & 0.776 & 0.703 & 0.407 & 0.338 & 0.171 & \textbf{0.995} & \textbf{0.873} & 0.654 & 0.366 & 0.333 & 0.169 \\
\gls{TSED} & AUROC & 0.977 & 0.952 & 0.879 & 0.672 & 0.607 & 0.599 & \textbf{0.995} & \textbf{0.972} & \textbf{0.895} & \textbf{0.684} & \textbf{0.624} & 0.597 \\
 & AP & 0.896 & 0.506 & 0.304 & 0.180 & 0.155 & 0.120 & \textbf{0.943} & \textbf{0.703} & \textbf{0.340} & \textbf{0.209} & \textbf{0.173} & 0.110 \\
CodeBERTScore & AUROC & 0.835 & 0.453 & 0.508 & 0.546 & 0.521 & 0.423 & \textbf{0.916} & \textbf{0.678} & \textbf{0.673} & \textbf{0.592} & \textbf{0.600} & \textbf{0.504} \\
 & AP & 0.772 & 0.088 & 0.128 & 0.212 & 0.197 & 0.085 & \textbf{0.811} & \textbf{0.135} & \textbf{0.179} & 0.206 & \textbf{0.226} & \textbf{0.098} \\
\midrule
FusionTop3 & AUROC & 0.999 & 0.959 & \underline{0.967} & 0.825 & 0.756 & \underline{0.739} & \textbf{1.000} & \underline{\textbf{0.991}} & \underline{\textbf{0.981}} & \textbf{0.838} & \textbf{0.793} & \underline{\textbf{0.772}} \\
 & AP & 0.990 & 0.764 & \underline{0.719} & 0.463 & 0.351 & 0.216 & \textbf{0.999} & \textbf{0.849} & \textbf{0.726} & 0.439 & \textbf{0.372} & \textbf{0.227} \\
\midrule
JPlag & AUROC & 0.993 & 0.963 & 0.887 & 0.739 & 0.650 & 0.514 & 0.993 & 0.963 & 0.887 & 0.739 & 0.650 & 0.514 \\
 & AP & 0.934 & 0.732 & 0.408 & 0.301 & 0.244 & 0.108 & 0.934 & 0.732 & 0.408 & 0.301 & 0.244 & 0.108 \\
Dolos & AUROC & 0.998 & \underline{0.983} & 0.938 & \underline{0.842} & \underline{0.790} & 0.690 & 0.998 & 0.983 & 0.938 & \underline{0.842} & 0.790 & 0.690 \\
 & AP & 0.991 & 0.842 & 0.709 & \underline{0.544} & \underline{0.402} & 0.207 & 0.991 & 0.842 & 0.709 & \underline{0.544} & 0.402 & 0.207 \\
\bottomrule
\end{tabular}
}
\end{table*}

Moving to the results per plagiarism level as shown in Table \ref{tab:pooled_level_auc_ap_raw_vs_pp}, with both raw and preprocessed code pairs, and both \gls{AUROC} and \gls{AP} values across plagiarism levels (L1-L6). Without preprocessing, CodeBLEU is best on L1, Dolos attains the highest \gls{AUROC} on L2 while CodeBLEU-syntax attains the highest \gls{AP}, FusionTop3 leads L3 in both \gls{AUROC} and \gls{AP}, Dolos is best on L4 and L5, and on L6 FusionTop3 gives the highest \gls{AUROC} while CrystalBLEU gives the highest \gls{AP}. With preprocessing, FusionTop3 leads \gls{AUROC} on L2, L3, and L6, CrystalBLEU leads both \gls{AUROC} and \gls{AP} on L5 and the \gls{AP} on L3 and L6, while Dolos remains the strongest method on L4. Overall, L1 and L2 are the easiest levels, whereas performance drops from L4 onward, particularly in \gls{AP}, leading to more false positives.

\section{Discussion}
This section covers the findings and their interpretation and implications, discusses the limitations, and mentions the threats to validity. 

\subsection{Findings}

The results presented in the previous section show consistent patterns across datasets, plagiarism levels, and evaluation metrics. In this section, we interpret these findings and examine what they imply about the effectiveness of \glspl{CEM} compared to established plagiarism detection tools. We focus on three main aspects: overall detection performance, behaviour across plagiarism levels, and the impact of preprocessing. The discussion connects the quantitative results to practical implications for \gls{SCPD} and highlights where performance remains strong and where limitations appear.

The results show that \glspl{CEM} can achieve comparable ranking performance to current established plagiarism detection tools. Several metrics (CodeBLEU, CrystalBLEU, and RUBY) had higher performance than JPlag. On the raw pooled setting, Dolos achieved the highest \gls{AUROC}, while CrystalBLEU was the strongest individual metric and FusionTop3 achieved the highest \gls{AP}. After preprocessing, FusionTop3 and CrystalBLEU surpassed Dolos on the pooled results. CrystalBLEU removes the common code and potentially the template, keeping the important part of the code to be compared, which makes it suitable for plagiarism detection and keeps it the strongest individual metric across the evaluated \glspl{CEM}. \gls{TSED} did not perform well on the task because it focuses only on structural similarity. Therefore, heavy lexical changes, such as reordering statements and adding wrappers, were not well-captured by this metric. CodeBERTScore failed to separate between plagiarised and non-plagiarised pairs because all pairs produced high similarity values. This aligns with the findings of previous work \cite{dristi2025analyzing}, \cite{nikiema2025small}. 

Preprocessing consistently improved the performance of several code metrics. Through investigation, we observe that preprocessing impacts the syntactic metrics and the embedding-based metric, CodeBERTScore. The n-grams are affected as they process the comments, and this introduces irrelevant or unnecessary tokens. So, removing comments improved metrics such as CrystalBLEU. In the components of CodeBLEU, we observe that the lexical n-gram components had a greater performance increase than the structure-based components with \gls{AST} and data-flow. The increase is the highest in CodeBERTScore, as preprocessing helps with the limited token length of that metric, as it reduces the overall token length. 

Detection performance depends on the plagiarism level. Both tools and most \glspl{CEM} effectively detected plagiarism across levels L1-L3. From L4 onwards, performance decreases across all methods. We observe that the scenarios that both metrics and tools frequently fail to detect are related to multiple code transformations and the creation of additional functions or classes. Therefore, they fail to detect more complex plagiarism scenarios. 

The exploratory fusion approach shows more consistent performance. It is the combination of three metrics, each of which captures different characteristics of source code. We observe that, on average, the fusion improves over the constituent individual metrics. This suggests that \glspl{CEM} can be combined with different similarity algorithms in the future to improve performance. 

Dataset characteristics also influence metric behaviour. IRPlag differs from ConPlag1 and ConPlag2 in its class distribution, containing 365 plagiarised and 95 non-plagiarised pairs, whereas the ConPlag datasets contain more non-plagiarised than plagiarised pairs. This imbalance corresponds to differences observed between \gls{ROC} and \gls{PR} behaviour. The results indicate that evaluation outcomes are influenced by dataset composition and class distribution.

Overall, the findings show that \glspl{CEM}, particularly when combined and supported by preprocessing, provide comparable performance to established plagiarism detection tools. 

\subsection{Answers to Research Questions}
We arrive at the following answers to the research questions:

\begin{itemize}

\item[\textbf{RQ1:}] 
Without preprocessing, the \glspl{CEM} lag behind Dolos, but have better ranking performance than JPlag. CrystalBLEU is the best-performing individual metric, as it removes the template or repetitive code. CodeBLEU and RUBY follow CrystalBLEU, with \gls{TSED} and CodeBERTScore performing worst among the evaluated metrics. FusionTop3 further narrows the gap to Dolos. 

\item[\textbf{RQ2:}] 
Preprocessing improved the performance measures for most \glspl{CEM}. This led FusionTop3 and CrystalBLEU to outperform Dolos on the pooled results and on ConPlag2, while Dolos remained best on ConPlag1. On IRPlag, CrystalBLEU achieved the highest \gls{AUROC} and CodeBLEU n-gram achieved the highest \gls{AP}. At the level-wise view, Dolos remained best on L4, whereas FusionTop3 and CrystalBLEU led most of the other challenging levels. 

\item[\textbf{RQ3:}] 
Without preprocessing and over the pooled dataset, the individual \glspl{CEM} remain slightly below Dolos in terms of \gls{AUROC}, although FusionTop3 nearly matches Dolos and achieves the highest \gls{AP}, and the strongest individual metrics remain above JPlag. With preprocessing, FusionTop3 and CrystalBLEU achieve better pooled \gls{AUROC} than both tools. Dolos still performs best on ConPlag1 and on L4.

\end{itemize}

\subsection{Code Evaluation Metrics Limitations}

CodeBERTScore has two main limitations, which are generating high similarity scores (>0.99) over plagiarised and non-plagiarised pairs, which makes it difficult to differentiate between them. The second limitation is related to the limited token length (<512), and code after that gets truncated. This aligns with the findings of \cite{nikiema2025small}. TSED focuses more on \glspl{AST} and structural similarity and struggles with heavy syntax modification and further logical modifications. CodeBLEU, CrystalBLEU, and RUBY, along with JPlag and Dolos, still find it hard to detect higher levels of similarity, such as L5 and L6. They lead to a high number of false positives, leading to lower \gls{AP} values. 

\subsection{Threats to Validity}
This work is subject to several threats to validity. Firstly, the plagiarism levels L1 to L6 in ConPlag were manually labelled, which may slightly affect the per-level details and analysis. Secondly, all \glspl{CEM} and plagiarism detection tools were used with default settings. No hyperparameter tuning or threshold optimisation was performed. Default configurations were used to reflect realistic usage scenarios. FusionTop3 was defined based on observed performance. Different parameters can lead to different results. Thirdly, the evaluation is conducted only on monolingual Java datasets. Therefore, this work does not consider other programming languages or cross-language plagiarism. Finally, JPlag was used initially during the annotation of the ConPlag datasets, which may introduce bias towards it during the evaluation of these datasets. 

Regarding the experimental work, we focused the comparative study on the ranking metrics. In future work, we should focus on retrieval and classification metrics. 

\subsection{Implications}

In educational settings, \glspl{CEM} are better suited for screening or filtering than for automatic judgment. Top \glspl{CEM} can rank potentially plagiarised pairs higher than non-plagiarised ones. However, false positives and weaker performance on complex cases (L4-L6) mean that final decisions should remain with the instructor. These methods can support academic integrity efforts, but they should not be used in isolation. 

\glspl{CEM} do not replace dedicated plagiarism detection tools, but they can complement them in specific cases. Based on the results, a simple workflow is recommended. First, apply preprocessing. Then, use a strong ranking method, such as CrystalBLEU, to flag suspicious pairs. Finally, confirm these cases using a dedicated plagiarism tool and manual inspection. 

From a research perspective, the results suggest that future work should focus on combining complementary similarity metrics rather than relying on a single method. The improved performance of FusionTop3 indicates that different representations capture distinct aspects of code similarity.

The consistent performance drop from L4 onwards highlights a key limitation of current approaches. Methods that better capture program behaviour and semantic equivalence are needed to address higher-level plagiarism. 

\section{Conclusion}

This paper investigated whether \glspl{CEM} can be adapted for \gls{SCPD} in software engineering education. Using two open-source level-labelled datasets, ConPlag and IRPlag, we evaluated five representative metrics and compared their performance against established plagiarism detection tools, JPlag and Dolos, using threshold-free ranking-based measures. The results showed that CrystalBLEU, CodeBLEU, and RUBY achieve competitive overall performance and outperform JPlag on the pooled raw setting, while Dolos retains the highest raw \gls{AUROC}. FusionTop3 nearly matches Dolos on \gls{AUROC} and achieves the highest raw \gls{AP}. The metrics performed strongly at lower plagiarism levels and remained competitive on L6, but performance dropped from L4 onward. Preprocessing improved most of the metrics, allowing FusionTop3 and CrystalBLEU to surpass Dolos on the pooled results, while Dolos remained strongest on ConPlag1 and on L4.

As the metrics and tools struggle with larger and heavier changes in structure or logic, we will aim to develop a new tool or system that combines \glspl{CEM} with complementary algorithms or approaches that focus more on semantics, leading to better detection on L5-L6 plagiarism levels. 

Future work will consider additional evaluation metrics for retrieval and ranking, hyperparameter tuning per tool and \gls{CEM}, additional programming languages, and cross-language plagiarism scenarios. Also, the area of AI-obfuscated or AI-generated code is another possible direction for future work. 

We provide the code at a public GitHub repository \url{https://github.com/FahadEbrahim/CEM-SCPD}, as well as some supplementary results and materials.

\bibliographystyle{ACM-Reference-Format}
\bibliography{software}

\appendix

\section{Bootstrapped Confidence Intervals and Paired Comparisons}

\subsection{Pooled Confidence Intervals}
\label{app:pooled_confidence_intervals}

Tables \ref{tab:ci_raw_pooled} and \ref{tab:ci_pp_pooled} report bootstrapped 95\% confidence intervals for AUROC and \gls{AP} in the pooled setting, for both raw and preprocessed code pairs.

\begin{table}[h]
\centering
\caption{Bootstrapped 95\% confidence intervals for pooled raw results.}
\label{tab:ci_raw_pooled}
\begin{tabular}{lcccc}
\toprule
Method & AUROC & 95\% CI & AP & 95\% CI \\
\midrule
FusionTop3 & 0.861 & [0.845, 0.878] & 0.845 & [0.824, 0.861] \\
CrystalBLEU & 0.850 & [0.832, 0.866] & 0.834 & [0.813, 0.854] \\
Dolos & 0.864 & [0.848, 0.879] & 0.842 & [0.823, 0.858] \\
CodeBLEU & 0.837 & [0.821, 0.854] & 0.817 & [0.795, 0.838] \\
RUBY & 0.822 & [0.804, 0.842] & 0.811 & [0.789, 0.832] \\
JPlag & 0.777 & [0.753, 0.798] & 0.762 & [0.736, 0.787] \\
\bottomrule
\end{tabular}
\end{table}

\begin{table}[h]
\centering
\caption{Bootstrapped 95\% confidence intervals for pooled preprocessed results.}
\label{tab:ci_pp_pooled}
\begin{tabular}{lcccc}
\toprule
Method & AUROC & 95\% CI & AP & 95\% CI \\
\midrule
FusionTop3 & 0.882 & [0.866, 0.897] & 0.862 & [0.841, 0.878] \\
CrystalBLEU & 0.879 & [0.862, 0.895] & 0.865 & [0.848, 0.885] \\
Dolos & 0.864 & [0.848, 0.879] & 0.842 & [0.823, 0.858] \\
CodeBLEU & 0.843 & [0.825, 0.860] & 0.822 & [0.801, 0.842] \\
RUBY & 0.839 & [0.821, 0.859] & 0.826 & [0.804, 0.846] \\
JPlag & 0.777 & [0.753, 0.798] & 0.762 & [0.736, 0.787] \\
\bottomrule
\end{tabular}
\end{table}

\subsection{Paired Bootstrap Differences}
\label{app:paired_bootstrap_differences}

We further computed paired bootstrap differences for AUROC in the pooled setting, focusing on the top-performing metrics against Dolos. For comparisons against JPlag, we report AUROC differences computed from the rounded values in Tables \ref{tab:ci_raw_pooled} and \ref{tab:ci_pp_pooled}.

In the raw setting, FusionTop3, CrystalBLEU, CodeBLEU, and RUBY all exceed JPlag, with $\Delta$AUROC values of +0.084, +0.073, +0.060, and +0.045, respectively. Relative to Dolos, FusionTop3 is statistically indistinguishable ($\Delta$AUROC = -0.0023, 95\% CI [-0.0155, 0.0104]), and CrystalBLEU is also not clearly separated from Dolos because its interval includes zero ($\Delta$AUROC = -0.0142, 95\% CI [-0.0285, 0.0001]). CodeBLEU and RUBY are lower than Dolos ($\Delta$AUROC = -0.0269, 95\% CI [-0.0433, -0.0106] and $\Delta$AUROC = -0.0421, 95\% CI [-0.0579, -0.0269]).

In the preprocessed setting, the same four metrics remain above JPlag: FusionTop3 ($\Delta$AUROC = +0.105), CrystalBLEU ($\Delta$AUROC = +0.102), CodeBLEU ($\Delta$AUROC = +0.066), and RUBY ($\Delta$AUROC = +0.062). Relative to Dolos, FusionTop3 and CrystalBLEU are higher ($\Delta$AUROC = +0.0179, 95\% CI [0.0052, 0.0308] and $\Delta$AUROC = +0.0154, 95\% CI [0.0028, 0.0287]), whereas CodeBLEU and RUBY remain lower ($\Delta$AUROC = -0.0209, 95\% CI [-0.0380, -0.0035] and $\Delta$AUROC = -0.0247, 95\% CI [-0.0409, -0.0087]).

\end{document}